\documentclass[journal]{IEEEtran}
\usepackage{cite}
\usepackage{amsmath,amssymb,amsfonts}
\usepackage{algorithmic}
\usepackage{graphicx}
\usepackage{textcomp}
\usepackage{xcolor}
\def\BibTeX{{\rm B\kern-.05em{\sc i\kern-.025em b}\kern-.08em
    T\kern-.1667em\lower.7ex\hbox{E}\kern-.125emX}}

\usepackage{float}
\usepackage{caption}

\usepackage[font=footnotesize]{subfig}
\usepackage{graphicx}
\usepackage{dblfloatfix}    
\usepackage[hidelinks]{hyperref}

\usepackage{enumerate}
\usepackage{enumitem}

\begin{document}

\title{A Case Study of Image Enhancement Algorithms' Effectiveness of Improving Neural Networks' Performance on Adverse Images}

\author{Jonathan Sanderson,
         Syed Rafay Hasan,~\IEEEmembership{Senior Member,~IEEE,}
 }
\maketitle

\begin{abstract}

    Neural Networks (NNs) have become indispensable for applications of Computer Vision (CV) and their use has been ever-growing.
    NNs are commonly trained for long periods of time on datasets like ImageNet and COCO that have been carefully created to represent common ``real-world" environments.
    When deployed in the field, such as applications of autonomous vehicles, NNs can encounter adverse scenarios that degrade performance.
    Using image enhancements algorithms to enhance images before being inferenced on a NN model poses an intriguing alternative to retraining, however, published literature on the effectiveness of this solution is scarce.
    To fill this knowledge gap, we provide a case study on two popular image enhancement algorithms, Histogram Equalization (HE) and Retinex (RX).
    We simulate four types of adverse scenarios an autonomous vehicle could encounter, dark, over exposed, foggy, and dark \& rainy weather conditions.
    We evaluate the effectiveness of HE and RX using several well established models:, Resnet, GoogleLeNet, YOLO, and a Vision Transformer.
    
\end{abstract}


\section{Introduction} \label{intro}

    Today, Computer Vision (CV) is prevalently used.
    Through the use of Convolutional Neural Networks (CNNs) CV is able to preform complicated tasks such as image classification, object detection, segmentation, etc.
    Recently we have seen a new type of architecture, encoder-decoder networks, gain interest in the field of CV.
    In order to train Neural Networks (NNs), large dataset have to be procured.
    Typically, these datasets are composed of images resembling the world we see.
    ImageNet and Common Objects in Context (COCO) are examples of datasets commonly used to train NNs.
    While these datasets do an excellent job representing environments that CNNs encounter when deployed in the real-world, they cannot account for every environment.
    An example of this could be a autonomous driving cars using a NN trained for the task of recognizing people, may perform well during the day, but may miss people at night, due to not being trained on dark images.
    The common approach to solving this problem is training the CNN using image augmentation, images that have been modified by an algorithm to further the domain theCNN is trained on.
    This is an effective approach but requires retraining a CNN, which is undesirable given most modern CNNs are quite large and take a significant amount of time and resources to (re)train.
    A desirable alternative is avoiding retraining by using commonly available trained CNN models and use image enhancement algorithms to map an adverse image to a similar domain the CNN was trained on.
    An example of this being, brightening a dark image before it is inferenced on.
    While designing image enhancements is an entire field of research of its own, there is surprising little work published on the effect of using known image enhancement algorithms to improve inference of CNNs.
    Furthermore, to the authors' knowledge, there is no work published on the effect of image enhancement algorithms to counter environments with non-ideal visibility for the purpose of increasing performance of CNNs or similar Neural Networks (NNs).
    Typically, the research field in enhancing images is centered around using CNNs to enhance images such as \cite{wang2017dilated} \cite{nagano2018srgan} \cite{murad2022colornet} \cite{li2019hdrnet}.
    There are also several works published on using CNNs to enhance images of adverse conditions, \cite{retinex-net} \cite{suarez2018deep} \cite{wang2020model}.
    Though using the solution of using CNNs to enhance images is quite effective, CNNs are much more computationally intensive than classical image enhancement algorithms.
    Edge devices such as autonomous vehicles typically have latency requirements, where using algorithms are preferable over using CNNs for image enhancement.
    While authors in \cite{image-ehancment-thesis} explores using image enhancements to improve CNNs' performance, they do not investigate images of adverse environments.
    To fill this lack in the knowledge base, we provide a case study for two well known image enhancement algorithms, Histogram Equalization (HE) and Retinex (RX), which according to \cite{image-enhancement-survey} make up 47\% of paper published exploring image enhancement methods.
    We test these algorithms with four adverse scenarios autonomous vehicles could encounter.
    The scenarios being: dark, overexposed, foggy, and rainy conditions in dark environment.
    We evaluate the enhancement algorithms listed previously across several well known NN models, Resnet \cite{resnet}, YOLO \cite{yolov1}, GoogleLeNet \cite{googlenet}, and a Vision Transformer (ViT) \cite{vision-transformer} \cite{vit-model}.

\section{Color Spaces and Image Enhancement Algorithms} \label{background}
    To aid the reader, we are providing some context to terms used in this work.
    Commonly, images have three subpixels per every pixel to represent color information, each subpixel being one channel of the red green blue (RGB) color space.
    The images described use the RGB color space to represent color data.
    Each channel is typically represented with, 8 bits, allowing each channel to have 256 unique brightness values, referred to as levels.
    HE algorithm's goal is to increase the contrast and center the mean of image brightness.
    For example, a dark image yields a histogram slanted to the left, a bright image yields in an image slanted to the right.
    When the HE algorithm is applied to an image, it remaps the darkest pixels in an image to level 0, and the brightest levels in the image to 255.
    It does this for all levels, spreading them out between levels 0 through 255
    The HE algorithm is designed for use with images with one channel (commonly referred to as black and white images), as the HE algorithm only adjusts the brightness of an image. 
    The HE algorithm can be extended to work with color images, by first converting an image to the Hue Saturation Value (HSV) color space, and applying HE to the V channel of the converted image.
    This allows the HE to effect only the brightness of color images, without effecting the color balance.
    Retinex (RX) is a quite a bit more complicated, as it is designed to work on color images.
    RX tries to mimic the human retina, hence the name ``Retina-x".
    It uses the Gaussian kernel function to estimate the correct brightness and color balance of multiple areas in an image using spacial locality \cite{retinex-algorithm}.
    Both HE and RX are designed to enhance an image, but use completely different approaches.

\section{Our Approach for Image Augmentation} \label{img-augments}

\begin{figure}[h]
         \centering
         
         \subfloat[Original\label{fig:effects:original}]{\includegraphics[width=0.12\textwidth]{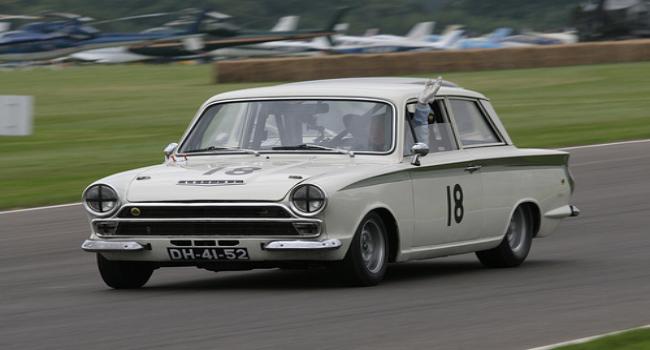}}
         
         \subfloat[Dark\label{fig:effects:dark}]{\includegraphics[width=0.12\textwidth]{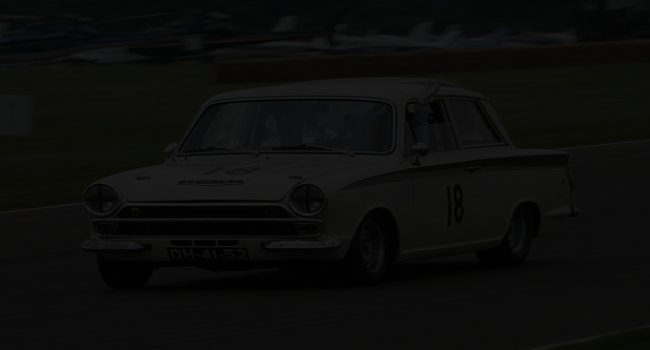}}
         \subfloat[Over Exposed \label{fig:effects:oe}]{\includegraphics[width=0.12\textwidth]{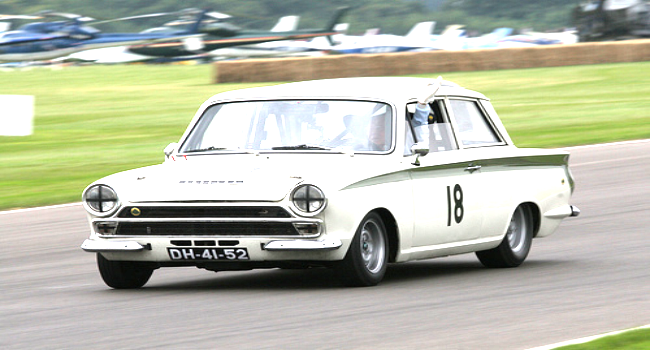}}
         \subfloat[Foggy\label{fig:effects:foggy}]{\includegraphics[width=0.12\textwidth]{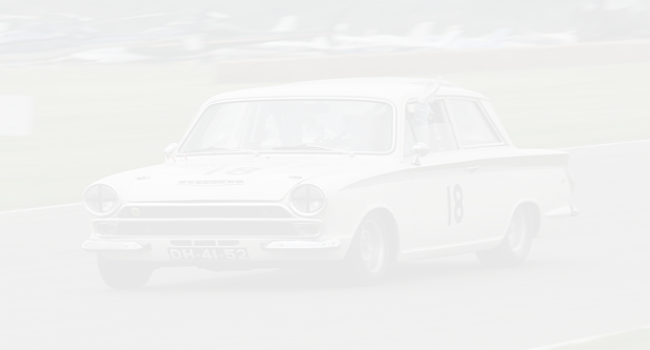}}
         \subfloat[Dark \& Rainy \label{fig:effects:dr}]{\includegraphics[width=0.12\textwidth]{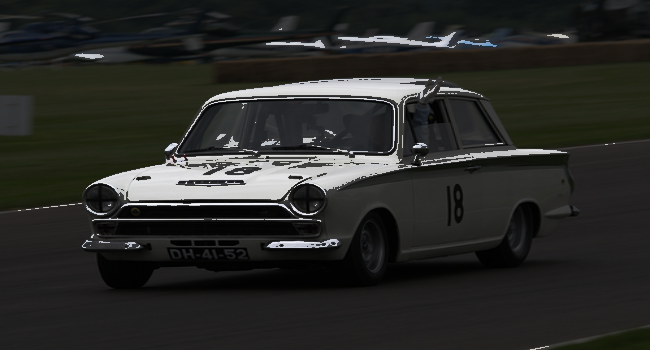}}
    
        \caption{Example of the four image augments applied to image \ref{fig:effects:original}}
        \label{fig:img_augments}
    \end{figure}

    As mentioned in before, HE and RX are families of algorithms.
    For implementation, we use OpenCV's \cite{opencv} HE function, and for RX we used the Single-Scale Retinex (SSR) function from a publicly available GitHub repository \cite{retinex_github}.
    The RX SSR algorithm requires a configuration value, in our preliminary testing a value of 100 is determined to be adequate through iterative testing.
    Image enhancement functions are not needed for images similar to the ones NN models have been trained on.
    Therefore, in order to evaluate the effectiveness of HE and RX in adverse environments, we need images of adverse environments.
    We have decided to simulate adverse environments, instead of using specialty datasets composed of images gathered from adverse environments.
    We reason there are two main advantages to this.
    First, we are easily able to achieve a wider variety of adverse conditions.
    Second, by using the validation sets of the datasets the models are trained on, we are able to have a ground truth for comparison.
    The process of modifying images to simulate adverse conditions, is referred to as \textit{augmentation} in this work.
    For this work, we are using four image augments to simulate conditions a self-driving vehicle may encounter.
    The four augments are: dark, to simulate driving at night, over exposure, to simulate driving in the evening or morning when traveling the direction of the sun, haze, for driving during foggy conditions, and dark \& rainy, to simulate driving in wet conditions at night.
    Bellow, we provide the formulas used to compute the image augments.
    We, apply the following equations to every pixel in an image. 
    $X$ represents a pixel of an image, $x$ represents a subpixel of $X$, $y$ represents the augmented pixel subpixel, $Y$ represents an augmented pixel, and $\hat{x}$ represents intermediate values.
    When an operation is applied to $x$, the operation is applied to every subpixel in $X$ implicitly.
    For dark augments, the following formula was used:
    \[ y = x / 8 \]
    This makes the image around an eighth as bright, as can be seen in Figure \ref{fig:effects:dark}.
    The equation used to simulate over exposure is as follows:
    \[ y = min(255, x \times 2) \]
    Where $min$ function, clips each subpixel in $x$ so that no value is greater than 255, thus simulating overexposure.
    The effect of this can be seen in Figure 1(c)
    For adding fog the equation is as follows:
    \begin{gather*}
        \hat{x} = x + (255 \times 10) \\
        y = \hat{x} \times 255 / image_{max}(\hat{x})
    \end{gather*}
    Where $image_{max}(\hat{x})$ is the maximum value of all calculated $\hat{x}$ in an image.
    $image_{max}$ is used to normalize the intermedia values to between 0 and 255.
    The result is the contrast is greatly decreased, and the average brightness of the image is greatly increased as can be seen in Figure 1(d).
    For dark and rainy augment the following equation is applied:
    \begin{gather*}
        \hat{x}_1 = (-x + 255) + (255 \times 2)     \\
        \hat{x}_2 = \hat{x}_1 \times 255 / image_{max}(\hat{x}_1)    \\
        \hat{x}_3 = -\hat{x}_2 + 255    \\
        Y = thres(X, \hat{X}_3, 255 \times 0.75)
    \end{gather*}
    The steps for the rain augment are similar to that of fog augment, but instead of the fog causing the image to become brighter, the image becomes darker. The $thresh$ function causes every pixel with any subpixel value greater than $255 \times 0.75$ to retain it original value.
    An example of this is shown in Figure \ref{fig:effects:dr}.

\section{Experimental Methodology} \label{exp-method}

    \subsection{Research Questions and Model Selection}
    We are seeking to answer the following research questions related to using image enhancement to counter adverse conditions in this case study:
    \begin{enumerate}[start=1,label={\bfseries RQ\arabic*)}, leftmargin=4\parindent]
        \item Are different CV tasks effected the same? (e.g. task of image classification vs. the task of object detection?) 
        \item What effect does changing model size have for the same model architecture?
        \item What effect does using a different model version have?
        \item Are different NN model architectures affected differently?
    \end{enumerate}
    To answer these questions, we are using a diverse set of NN models to see the effect of using image enhancements.
    The models used: Resnet18 \cite{resnet18}, Resnet50 \cite{resnet50}, Resnet101 \cite{resnet101}, YOLOv5n \cite{yolov5n}, YOLOv5m \cite{yolov5m}, YOLOv8n \cite{yolov8n}, YOLOv8m \cite{yolov8m}, GoogleLeNet \cite{googlenet-model} and a vision transformer (ViT) \cite{vit-model}.
    To answer RQ1 we are using several image classification model, Resnet, GoogleLeNet, and ViT, and YOLO a well established object detection model.
    To address RQ2 we are using three sizes of Resnet, Resnet18 having 11.6 million weights, Resnet50 having 25.5 million weights, and Resnet101 having 44.5 million weights.
    For RQ3, we are using two different sizes of two different versions of YOLO.
    The sizes being nano and medium, denoted `n' and `m', and version v5 and v8.
    We average the two sizes of each version of YOLO to better capture the 
    To answer RQ4 we are using Resnet (the average of Resnet 18, 50, and 101), GoogleLeNet, and ViT.
    Resnet and GoogLeNet are both CNNs that use different approaches to feature extraction, ViTs on the other hand use and an attention mechanism to extract information from images.

    \subsection{Testing Setup}
\begin{figure}[h]
        \centering
        \includegraphics[width=0.495\textwidth]{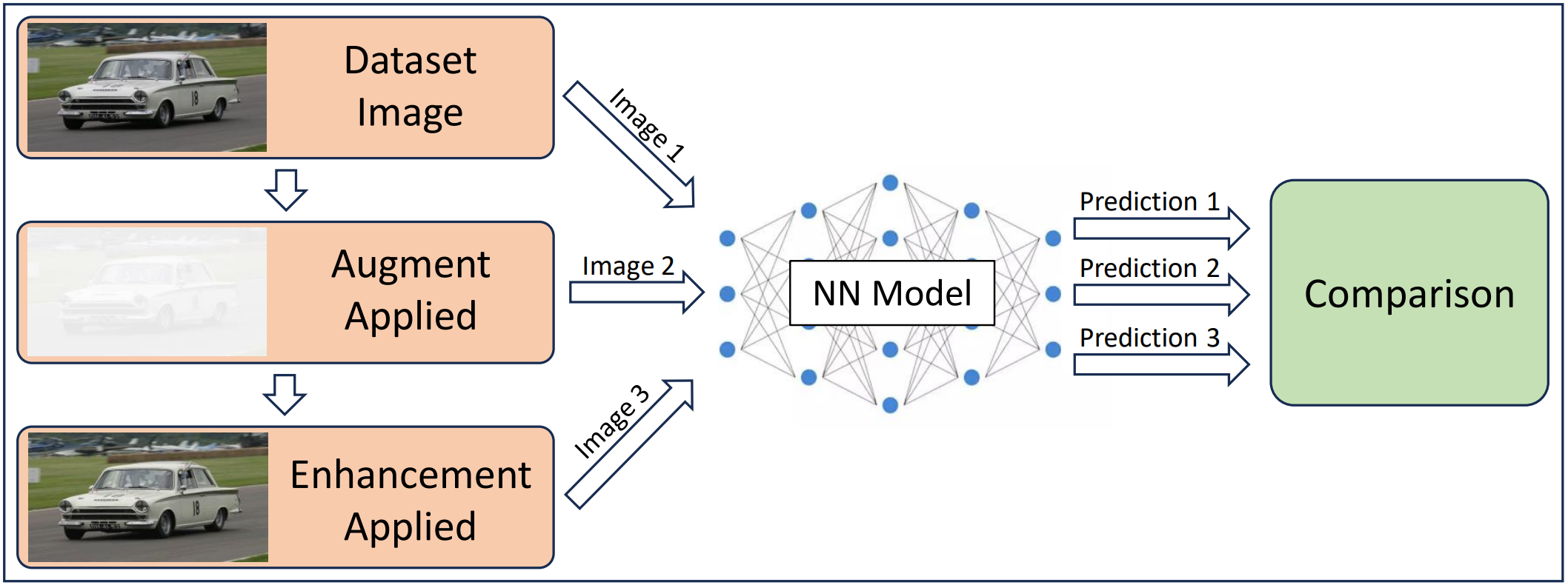}
        \caption{Visualization of testing methodology}
        \label{fig:methodology}
    \end{figure}

    What we are most focused on is doing a relative comparison of using image enhancement functions.
    That being, to observe the change in accuracy of NN model, before and after an image enhancement is applied to an image.
    So for each image augment, we test a NN model with and without HE and RX applied.
    We also conduct two ablation tests.
    These ablation tests being to observe the effect of using HE an RX without an image augment, e.g. image augment equation is $y = x$.
    A general overview of testing methodology is shown in Figure \ref{fig:methodology}.
    \subsection{Datasets, Preprocessing, and Metrics}
    We are using two validation datasets for our testing.
    The datasets used depends on the CV task.
    All image classifiers used, Resnet, GoogleLeNet, ViT, have been trained on the ImageNet dataset, so the Imagenet validation set is used to evaluate these models.
    For object detection, YOLO was trained on the COCO2017 training set, therefore the COCO2017 validation set is used to evaluate YOLO.
    For the preprocessing of the validation datasets, the only preprocessing performed is resizing of the images.
    This is not standard practice, as models are often trained with images that have been had their mean and standard deviation adjusted, among other transformations.
    We found in preliminary testing that the recommended preprocessing steps can interfere with our test setup.
    We also found that not using the recommended preprocessing steps marginally affects the baseline accuracy of a NN model.
    We use three different metrics for analysis.
    For evaluating the effect of image augments and image enhancements, we use the mean and standard deviation.
    For evaluating model performance, we used and average of Top-1 and Top-5 accuracy for the task of image classification and, mAP 50:95 for the task of object detection.

\section{Results}   \label{results}

\begin{figure}[h]
         \centering
    
         \subfloat[Dark + HE\label{fig:he:dark}]{\includegraphics[width=0.12\textwidth]{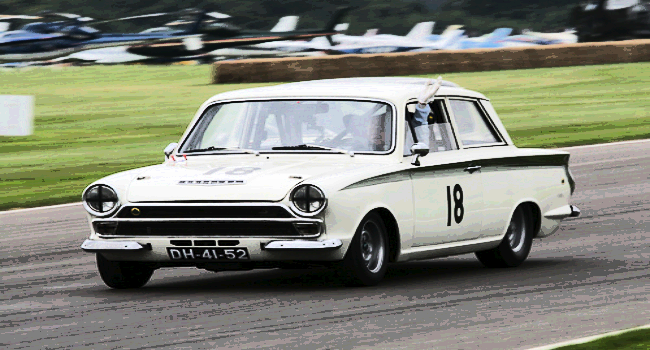}}
         \subfloat[Over Exposed + HE\label{fig:he:oe}]{\includegraphics[width=0.12\textwidth]{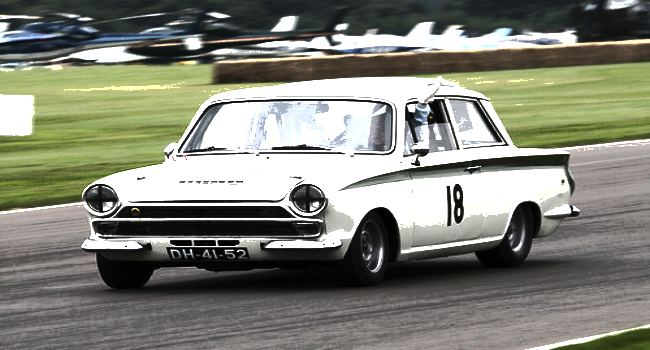}}
         \subfloat[Foggy + HE\label{fig:he:foggy}]{\includegraphics[width=0.12\textwidth]{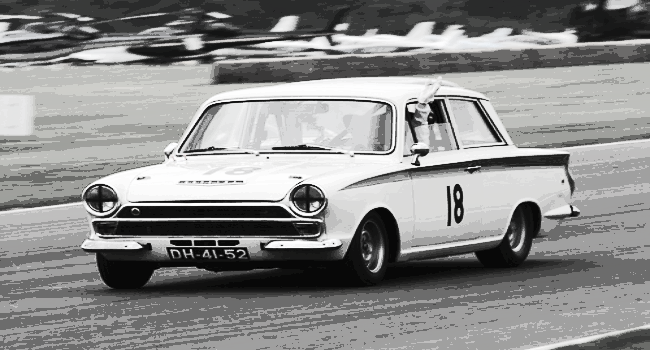}}
         \subfloat[Dark \& Rainy + HE\label{fig:he:dr}]{\includegraphics[width=0.12\textwidth]{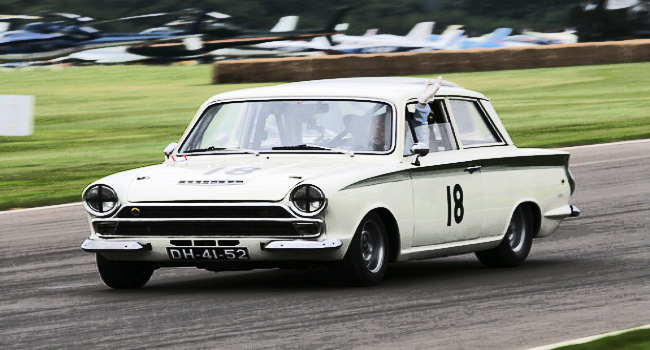}}

        \caption{Images after Histogram Equalization (HE) is applied}
        \label{fig:he}
    \end{figure}

\begin{figure}[h]
         \centering
    
         \subfloat[Dark + RX\label{fig:rx:dark}]{\includegraphics[width=0.12\textwidth]{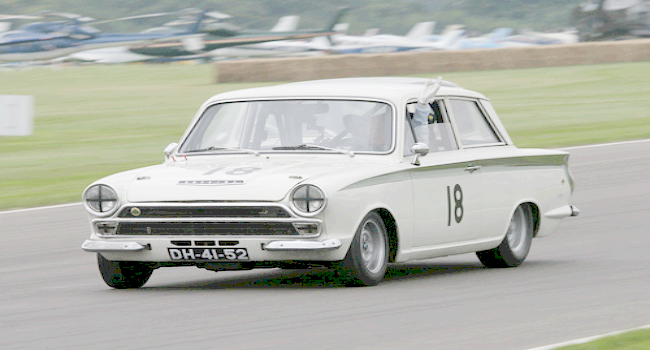}}
         \subfloat[Over Exposed + RX\label{fig:rx:oe}]{\includegraphics[width=0.12\textwidth]{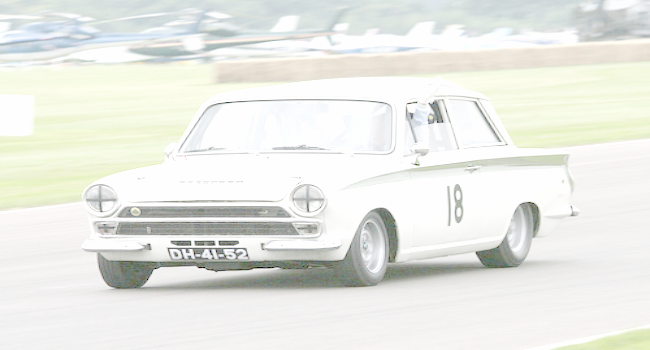}}
         \subfloat[Foggy + RX\label{fig:rx:foggy}]{\includegraphics[width=0.12\textwidth]{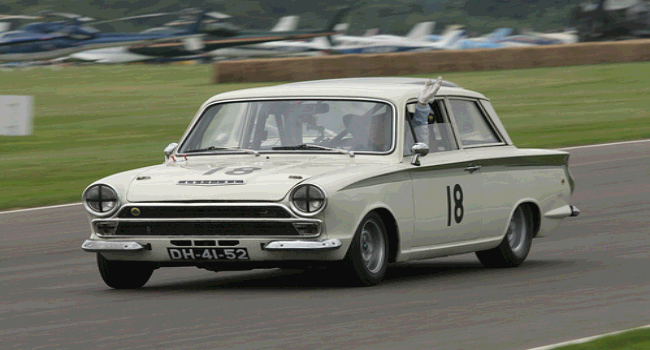}}
         \subfloat[Dark \& Rainy + RX\label{fig:rx:dr}]{\includegraphics[width=0.12\textwidth]{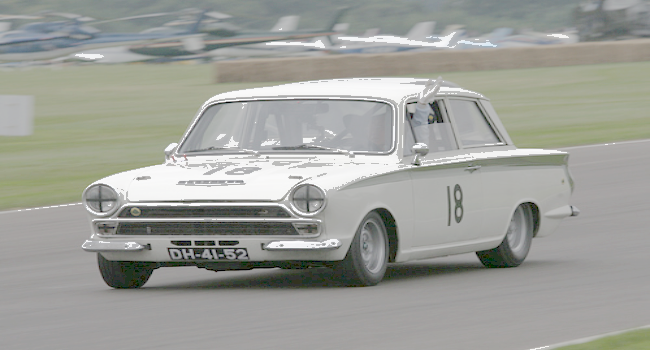}}

        \caption{Images after Retinex (RX) is applied}
        \label{fig:rx}
    \end{figure}

\begin{figure}[h]
        \centering
        \includegraphics[width=0.49\textwidth]{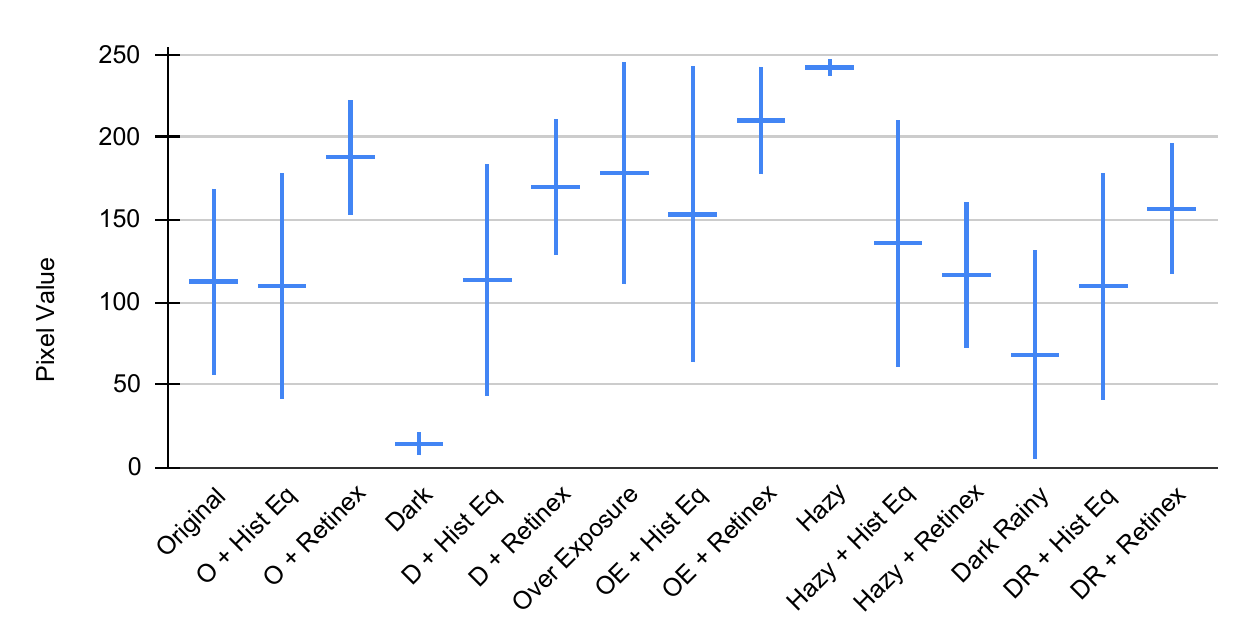}
        \caption{Comparison of mean and standard deviation of images before and after being enhanced. The blue horizontal lines represent the mean and the vertical lines depict one standard deviation above and below the mean}
        \label{fig:mean_std}
    \end{figure}

\begin{figure}[h]
        \centering
        \includegraphics[width=0.49\textwidth]{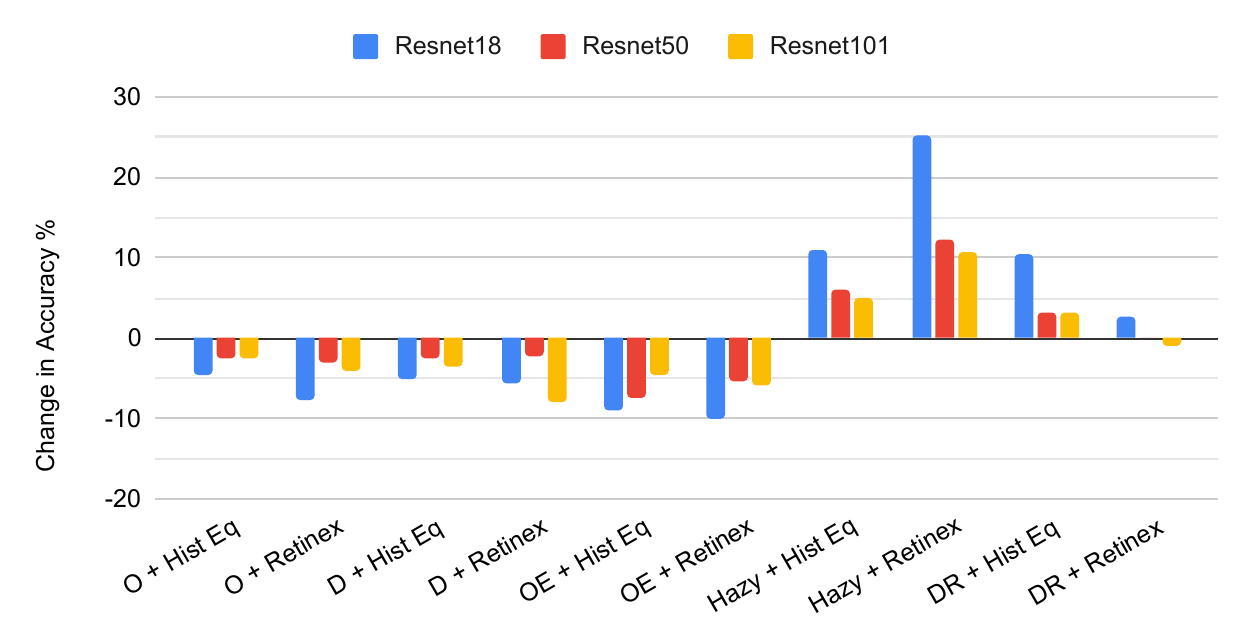}
        \caption{Change in accuracy across different sizes of Resnet50.}
        \label{fig:resnet-improvment}
    \end{figure}

\begin{figure}[h]
        \centering
        \includegraphics[width=0.49\textwidth]{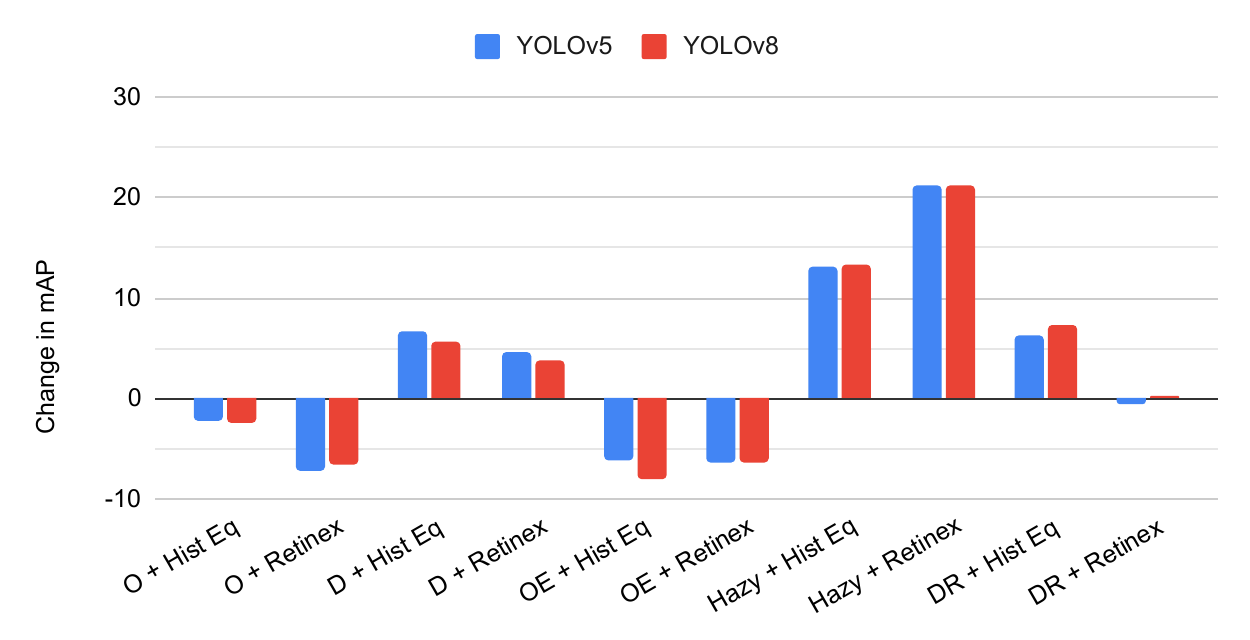}
        \caption{Change in mAP of Yolov5 and Yolov8}
        \label{fig:yolo-improvement}
    \end{figure}

\begin{figure}[h]
        \centering
        \includegraphics[width=0.49\textwidth]{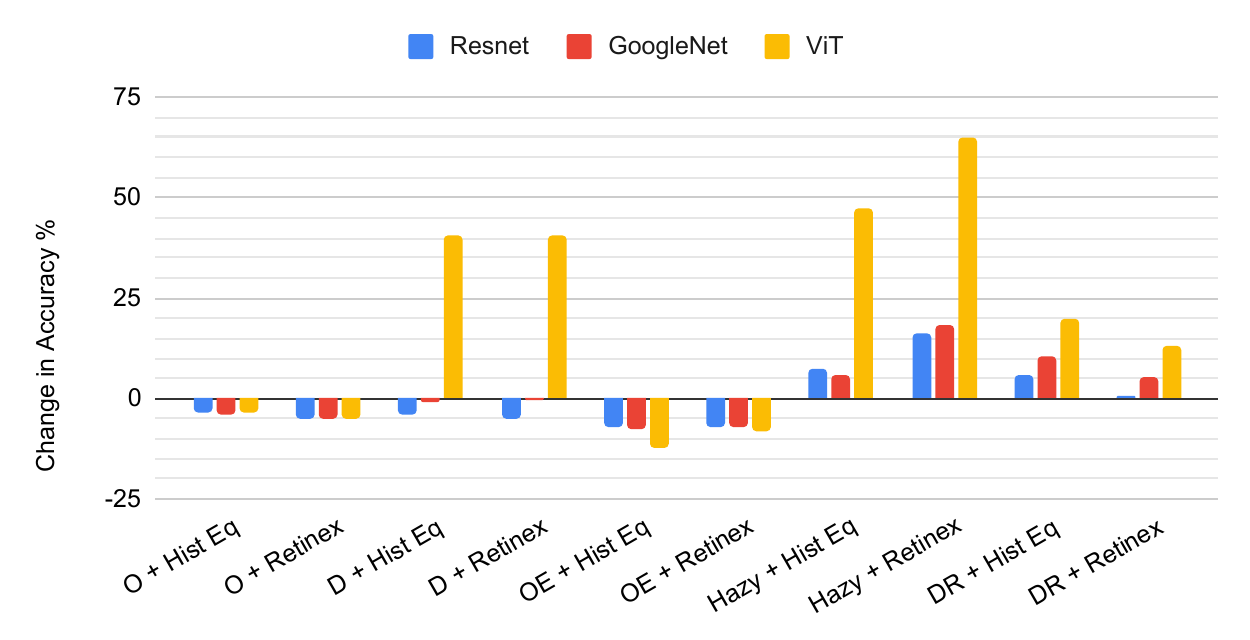}
        \caption{Change in accuracy across three different types of image classifiers: Resnet, GoogleLeNet, and ViT}
        \label{fig:3models-image-improvement}
    \end{figure}

    In Figure \ref{fig:he} an example of applying HE augmented images (see Figure \ref{fig:img_augments} for augmented images) is shown.
    The effect of applying RX to the augmented images is shown in Figure \ref{fig:rx}.
    Since the mean and standard deviation results for ImageNet and COCO2017 validation datasets are similar, therefore the results have been averaged together, shown in Figure \ref{fig:mean_std}.
    Figures \ref{fig:resnet-improvment}, \ref{fig:yolo-improvement} and \ref{fig:3models-image-improvement} show the change in accuracy and mAP for the evaluated NN models.
    These figures show change before and after an image enhancement is applied.
    A value on the vertical axis represent no change, a positive value indicates improvement, and negative values indicate a degradation in a NN model's performance.
    A note of clarity: If a model has an accuracy of 70\% before image enhancement is applied and has an accuracy of 85\% after image enhancement is applied, it will be shown as an increase of 15\% in figures \ref{fig:resnet-improvment} and \ref{fig:3models-image-improvement}.
    This should not be interpreted as the model is performing 15\% better.

\section{Discussion}

In the Experimental Methodology section, we sought to answer four questions.
Comparing Figures \ref{fig:yolo-improvement} and \ref{fig:resnet-improvment}, we can deduce there is no major difference between image classification and object detection task, for over exposed, hazy, and dark \& rainy image augments.
For the dark augment, it seems that the task of object detection is especially sensitive to low light environments, as HE and RX are both able to increase the mAP of YOLO.
Investigating Figure \ref{fig:resnet-improvment} reveals model size does play a role.
Resnet18 is more sensitive to HE and RX than Resnet50 and Resnet101 are.
However, looking at when RX is applied to dark images, Resnet101 decreases in accuracy more than Resnet18 and Resnet50.
In general, while model size does impact its sensitivity to image enhancements, especially for small models, it is not a linear correlation.
Looking at Figure \ref{fig:yolo-improvement} we can see there is minimal difference between YOLOv5 and YOLOv8, so changing the model version has minimal effect.
For the last question, Figure \ref{fig:3models-image-improvement} provides clear indication that image enhancement can have a wide-ranging effectiveness on different NN architectures.
The two CNN models, Resnet and GoogleLeNet react similarly to most image enhancements, this is not true for ViT.
These figures show that CNN results does become better if it is used in conjunction with image enhancement, especially in the cases of hazy conditions and dark with rainy conditions. We also found out that over exposed images (e.g. snowy conditions during daytime) does not affect the CNN performance significantly, hence utilization of image enhancement is not needed for this case (rather it diminishes the CNN performance). 

\section{Conclusion} \label{conculsion}
In this paper we have explored the effect of using two well established image enhancement algorithms, Histogram Equalization (HE) and Retinex (RX), on Neural Networks (NNs) inferencing adverse images.
We developed image augments to simulate four adverse scenarios that an autonomous vehicle could encounter, dark, over exposed, foggy, and dark \& rainy.
We evaluated multiple aspects of how effective HE and RX are at enhancing the four types of adverse images using several well known NNs, Resnet, GoogleLeNet, YOLO and ViT.
We found that for optimal results, the algorithm used must be intelligently chosen.
To aid researcher further explore this topic, we will provide our source code on a public website.
 
\newpage

\bibliographystyle{ieeetr}
\bibliography{refs}


\end{document}